\documentstyle[iopconf1]{article}
\input axodraw.sty

\def\lsim{\:\raisebox{-0.5ex}{$\stackrel{\textstyle<}{\sim}$}\:}

\newcommand{\newc}{\newcommand}
\newc{\pbi}{pb$^{-1}$}
\newc{\ie}{{\it i.e.} }
\newc{\ti}{\tilde}
\newc{\ra}{\rightarrow}
\newc{\ee}{$e^+e^-$\ }
\newc{\mm}{$\mu^+\mu^-$}
\newc{\taus}{$\tau^+\tau^-$}
\newc{\uu}{$u\bar{u}$\ }
\newc{\eeee}{$e^+e^-\ra e^+e^-$\ }
\newc{\eemm}{$e^+e^-\ra \mu^+\mu^-$\ }
\newc{\eett}{$e^+e^-\ra \tau^+\tau^-$\ }
\def\Rs{R \hspace{-0.38em}/\;}
\newc{\beq}{\begin{eqnarray}}
\newc{\eeq}{\end{eqnarray}}
\newc{\dqu}{\delta_{qu}}
\newc{\dqd}{\delta_{qd}}
\newc{\non}{\nonumber}
\newc{\noi}{\noindent}

\def\ib#1,#2,#3{       {\it ibid.\/ }{\bf #1} (19#2) #3}
\def\ap#1,#2,#3{       {\it Ann.~Phys.~(NY)\/ }{\bf #1} (19#2) #3}
\def\ijmp#1,#2,#3{     {\it Int.\ J.~Mod.\ Phys.\/ } {\bf A#1} (19#2) #3}
\def\mpla#1,#2,#3 {     {\it Mod.~Phys.~Lett.\/ } {\bf A#1} (19#2) #3}
\def\npb#1,#2,#3{       {\it Nucl.\ Phys.\/ }{\bf B#1} (19#2) #3}
\def\npps#1,#2,#3{     {\it Nucl.\ Phys.~B (Proc.~Suppl.)\/ }{\bf B#1}
                             (19#2) #3}
\def\plb#1,#2,#3{      {\it Phys.\ Lett.\/ }{\bf B#1} (19#2) #3}
\def\pr#1,#2,#3{       {\it Phys.\ Rev.\/ }{\bf #1} (19#2) #3}
\def\prd#1,#2,#3{      {\it Phys.\ Rev.\/ }{\bf D#1} (19#2) #3}
\def\prep#1,#2,#3{     {\it Phys.\ Rep.\/ }{\bf #1} (19#2) #3}
\def\prl#1,#2,#3{      {\it Phys.\ Rev.\ Lett.\/ }{\bf #1} (19#2) #3}
\def\pro#1,#2,#3{      {\it Prog.~Theor.\ Phys.\/ }{\bf #1} (19#2) #3}
\def\rmp#1,#2,#3{      {\it Rev.~Mod.~Phys.\/ }{\bf #1} (19#2) #3}
\def\sp#1,#2,#3{       {\it Sov.~Phys.~Usp.\/ }{\bf #1} (19#2) #3}
\def\zpc#1,#2,#3{      {\it Z.~Phys.\/ }{\bf C#1} (19#2) #3}
\def\appb#1,#2,#3{     {\it Acta Phys.\ Polon.\/ }{\bf B#1} (19#2) #3}

\begin{document}
\begin{flushright} DESY 97-145\\{\tt hep-ph/9708490} \end{flushright}
\title{Sleptons at LEP2 and Tevatron in $R$-Parity Violating
  SUSY\footnote{Talk presented at ``Beyond the Desert 97 --
    Accelerator and Non-Accelerator Approaches'', Ringberg Castle,
    Germany, 8-14 June 1997, to appear in the proceedings.}
}

\author{Jan Kalinowski\footnote{E-mail:
    Jan.Kalinowski@fuw.edu.pl.} \footnote{Supported in part by the
    EU Grant CIPD-CT94-0016.}}

\affil{Deutsches Elektronen-Synchrotron DESY, D-22607
  Hamburg\\[1.1ex] 
 Institute of Theoretical Physics, Warsaw University,
  PL-00681 Warsaw   }

\beginabstract
In supersymmetric theories with $R$-parity breaking, sleptons could be
produced singly in $e^+e^-$ collisions at LEP2 and in
$q\bar{q}$ annihilation at the Tevatron through interactions in
which two quark or two lepton fields are coupled to a slepton field.
At LEP they could manifest themselves in Bhabha scattering, 
and in the annihilation to $\mu^+\mu^-$, 
\taus, and $q\bar{q}$ pairs.  The formation of sneutrinos, $e^+e^-\ra
\ti{\nu}$, and their signals for a mass within the reach of this
machine, is an exciting speculation which can be investigated in the
coming LEP2 runs with energies close to $\sqrt{s}=200$ GeV. At the
Tevatron the sleptons can be searched for as resonances in
$p\bar{p}\rightarrow \tilde{\nu} \rightarrow \ell^+\ell^-$ and
$\tilde{\ell}\rightarrow \ell\nu $ final states. Existing LEP2 and
Tevatron data can be exploited to derive bounds on the Yukawa
couplings of sleptons to quark and lepton pairs.
\endabstract

\section{Introduction}
In the usual formulation, the minimal supersymmetric extension (MSSM)
of the Standard Model (SM) is defined by the superpotential which has
the form
\begin{equation}
  W_R=Y_{ij}^e L_iH_1 E^c_j + Y_{ij}^d Q_i H_1D^c_j
  +Y_{ij}^uQ_iH_2U^c_j +  \mu H_1 H_2 \label{Rcons}
\end{equation} 
The indices $i,j$ denote the generations, and a summation is
understood, $Y^f_{ij}$ are Yukawa couplings and $\mu$ is the Higgs
mixing mass parameter.  The standard notation is used in eq.\ 
(\ref{Rcons}) for the left-handed doublets of leptons ($L$) and quarks
($Q$), the right-handed singlets of charged leptons ($E$) and
down-type quarks ($D$), and for the Higgs doublets which couple to the
down ($H_1$) and up quarks ($H_2$). 

The interaction lagrangian derived from $W_R$ contains terms in which
the supersymmetric partners appear only in pairs. As a result,
superpartners can be produced only in pairs in collisions and decays
of particles, and the lightest supersymmetric particle (LSP) is stable. This
feature is traced to a discrete multiplicative symmetry of the
superpotential, $R$-parity, which can be defined as \cite{FF}
\begin{equation}
             R_p=(-1)^{3B+L+2S}
\end{equation} 
with $S$ denoting the spin of the particle: all Higgs particles and SM
fermions and bosons have $R_p=+1$, and their superpartners
have $R_p=-1$.

However, the gauge and Lorentz symmetry also allows for additional
terms in the superpotential which break the $R$-parity \cite{WSY}
\begin{equation}
  W_{\Rs}=\lambda_{ijk}L_iL_jE^c_k + \lambda'_{ijk}L_iQ_jD^c_k
  +\lambda''_{ijk} U^c_iD^c_jD^c_k\label{Rviol} +
  \epsilon_i L_i H_2 
\end{equation} 
with additional Yukawa couplings $\lambda$, $\lambda'$, $\lambda''$
and dimensionful mass parameters $\epsilon$. If these terms are
present, the model has distinct features: superpartners can be
produced singly and the LSP is not stable.  Because of
anti-commutativity of the superfields, $\lambda_{ijk}$ can be chosen to
be non-vanishing only for $i < j$ and $\lambda''_{ijk}$ for $j<k$.
Therefore for three generations of fermions, $W_{\Rs}$ contains
additional 48 new parameters beyond those in eq.~(\ref{Rcons}). Note
that at least two different generations are coupled in the purely
leptonic or purely hadronic operators.

The couplings $\lambda$, $\lambda'$ and $\epsilon$ violate lepton
number ($L$), whereas $\lambda''$ couplings violate baryon number
($B$), and thus can possibly lead to fast proton decay.  In the usual
formulation of the MSSM they are forbidden by $R$-parity.  However,
there is no theoretical motivation for imposing $R$-parity.  From the
grand unification and string theory points of view, both types of
models, $R_p$ conserving or violating, have been constructed with no
preference for either of the two \cite{D1}.  Since they lead to very
different phenomenology, both models should be searched for
experimentally.  The usual formulation of MSSM with $R_p$-conservation
has been extensively studied phenomenologically and experimentally.
Only recently the $R_p$-violating formulation has received more
attention as providing potentially favored solutions to some
experimental observations (if not fluctuations), like Aleph 4-jet
events \cite{cglw}, and HERA high $Q^2$ events \cite{HERAsq}.

$R_p$ conservation guarantees the stability of the proton by removing
all $\lambda$, $\lambda'$, $\lambda''$ and $\epsilon$ couplings.
However other discrete symmetries can allow for a stable proton and
$R$-parity violation at the same time. For example,
baryon-parity (defined as $-1$ for quarks, and $+1$ for leptons and
Higgs bosons) implies $\lambda''=0$. In this case only lepton number
is broken, which suffices to ensure proton stability, and to explain
experimental observations mentioned above. In addition, lepton-number
violating operators can provide new ways to generate neutrino masses. 
Since in supersymmetry lepton and $H_1$ fields have
the same quantum numbers, the last term in eq.~(\ref{Rviol}) can be
rotated away\footnote{For the discussion of $\epsilon_i L_i H_2$ term,
  see talk by J. Valle \cite{Valle}.} by a redefinition of $L_i$ and
$H_1$. Therefore we will consider below only the most general trilinear
terms in eq.~(\ref{Rviol}) that violate $L$ but conserve $B$.

In four-component Dirac notation, the $\lambda$ and $\lambda'$ part of
the Yukawa interactions has the following form: 
\begin{eqnarray} {\cal
  L}_{\Rs}&=&\lambda_{ijk}\left[ \ti{\nu}^j_L\bar{e}^k_Re^i_L
+\overline{\ti{e}}^k_R (\bar{e}^i_L)^c\nu^j_L
+\ti{e}^i_L\bar{e}^k_R\nu^j_L\right. \non \\
&&\left. \mbox{~~~~~~~~~}
-\ti{\nu}^i_L\bar{e}^k_Re^j_L-\overline{\ti{e}}^k_R (\bar{e}^j_L)^c\nu^i_L
-\ti{e}^j_L\bar{e}^k_R\nu^i_L \right] + h.c.  \nonumber \\ 
&+&\lambda'_{ijk}\left[( \ti{u}^j_L\bar{d}^k_Re^i_L
+\overline{\ti{d}}^k_R(\bar{e}^i_L)^cu^j_L +\ti{e}^i_L\bar{d}^k_Ru^j_L
)\right. \nonumber \\ &&\left.
\mbox{~~~~~~~~~}-(\ti{\nu}^i_L\bar{d}^k_Rd^j_L+\ti{d}^j_L\bar{d}^k_R\nu^i_L
+\overline{\ti{d}}^k_R(\bar{\nu}^i_L)^cd^j_L) \right] + h.c.\mbox{~~~}
\label{effl}
\end{eqnarray} 
where $u_i$ and $d_i$ stand for $u$- and $d$-type quarks, $e_i$ and
$\nu_i$ denote the charged leptons and neutrinos of the $i$-th
generation, respectively; the scalar partners are denoted by a tilde.
In the $\lambda'$ terms, the up (s)quarks in the first parentheses
and/or down (s)quarks in the second may be Cabibbo rotated in the
mass-eigenstate basis.  As we will discuss mainly sneutrino induced
processes, we will assume the basis in which only the up sector is
mixed, $i.e.$ the $NDD^c$ is diagonal. We will comment on choosing a
different basis, where relevant for our discussion.

This scenario can be explored in various processes.  Actually, the
possibility of some $\lambda$ and $\lambda'$ couplings to be
simultaneously non-zero opens a plethora of interesting processes at
current and future colliders. Since from low-energy experiments the
Yukawa couplings are expected to be small, indirect effects due to
$t/u$-channel exchanges of sfermions in collisions of leptons and
hadrons can be difficult to observe.  However, the direct formation of
sfermion resonances in the $s$-channel processes can produce
measurable effects.  For example, squarks could be produced as
$s$-channel resonances in lepton-hadron collisions at HERA.  In fact,
recent high $Q^2$, high $x$ events at HERA have been analyzed in this
context; higher statistics however is needed to draw definite
conclusions.  Sleptons on the other hand could be produced as
$s$-channel resonances in lepton-lepton and hadron-hadron collisions,
and could decay to leptonic or hadronic final states in addition to
$R$-parity conserving modes.

In this talk we will consider the possible effects of $s$-channel
slepton resonance production in $e^+e^-$ collisions
\begin{eqnarray}
&&e^+e^- \rightarrow \tilde{\nu}\rightarrow \ell^+\ell^- \\
&&e^+e^- \rightarrow \tilde{\nu}\rightarrow q\bar{q}
\end{eqnarray}
and in $p\bar{p}$ collisions 
\begin{eqnarray}
&& p\bar{p} \rightarrow \tilde{\nu}\rightarrow \ell^+\ell^- \\
&& p\bar{p} \rightarrow \tilde{\ell}^+\rightarrow \ell^+\nu  
\end{eqnarray}
We will not consider hadronic final states of sleptons produced in
$p\bar{p}$ collisions as they can be difficult to analyze
experimentally in the hadronic environment. The results presented here
have been obtained in collaboration with H. Spiesberger, R. R\"uckl
and P. Zerwas \cite{snu,tev}.

Note that since in SUSY GUT scenarios sleptons are generally expected
to be lighter than squarks, sleptons may show up at LEP2 and/or
Tevatron even if squarks are beyond the kinematical reach of HERA.

\section{Sfermion Exchanges in $f\bar{f}'\rightarrow F\bar{F}'$ Processes}
Before discussing specific reactions let us consider a generic
two-body process $f\bar{f}'\rightarrow F\bar{F}'$. In the Standard
Model it can proceed via $s$- and/or $t$-channel gauge boson exchange
($\gamma$, $Z$, or $W$; for light fermions the Higgs boson exchange is
negligible), as shown in Fig.~1.
Sfermions can contribute via $s$-, $t$-, and/or $u$-channel exchange
processes, Fig.~1. The differential cross section in the $f\bar{f}'$ 
rest frame 
can be written most transparently in terms of helicity amplitudes
\begin{eqnarray}
&&\frac{\mbox{d}\sigma}{\mbox{d}\cos\theta} (f\bar{f}' \ra F\bar{F}') 
= A_c
\frac{\pi\alpha^2s}{8}
\Bigl\{ 4\left[|f^t_{LL}|^2+|f^t_{RR}|^2\right]\non \\
&& \mbox{~~~~~~~~~} +  
 (1-\cos\theta)^2\left[|f^s_{LL}|^2+|f^s_{RR}|^2\right]
\non\\
&&\mbox{~~~~~~~~~} + (1+\cos\theta)^2 
\left[ |f^s_{LR}|^2 + |f^s_{RL}|^2 +
  |f^t_{LR}|^2 + |f^t_{RL}|^2 
\right.\non \\ 
&&\left. \mbox{~~~~~~~~~~~~~~~~~~~~~~~~~~}
+ 2\mbox{Re}(f^{s\;*}_{LR}\,f^t_{LR} + 
     f^{s\;*}_{RL}\,f^t_{RL}) \right] 
   \Bigr\}
\label{dsigdcos}
\end{eqnarray}
where $A_c$ is the appropriate color factor. 
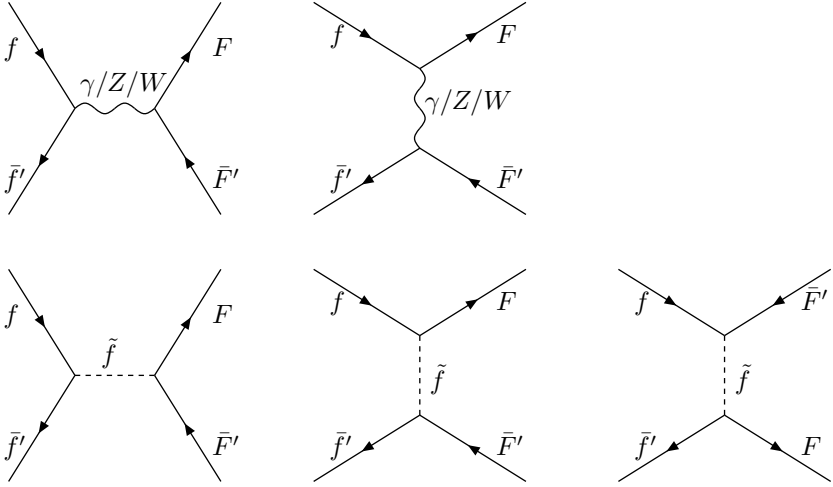
\begin{figure}[htbp]
%
\begin{picture}(100,100)(0,0)
\ArrowLine(10,90)(35,50)
\ArrowLine(35,50)(10,10)
\Photon(35,50)(65,50){2}{2}
\ArrowLine(90,10)(65,50)
\ArrowLine(65,50)(90,90)
\put(8,20){$\bar{f}'$}
\put(8,70){$f$}
\put(37,57){$\gamma/Z/W$}
\put(87,20){$\bar{F}'$}
\put(87,70){$F$}
\end{picture}
%
\begin{picture}(100,100)(-12,0)
\ArrowLine(10,90)(50,65)
\ArrowLine(50,65)(90,90)
\Photon(50,65)(50,35){2}{2}
\ArrowLine(90,10)(50,35)
\ArrowLine(50,35)(10,10)
\put(16,20){$\bar{f}'$}
\put(16,75){$f$}
\put(52,50){$\gamma/Z/W$}
\put(79,20){$\bar{F}'$}
\put(79,75){$F$}
\end{picture}
\\
%
\begin{picture}(100,100)(0,0)
\ArrowLine(10,90)(35,50)
\ArrowLine(35,50)(10,10)
\DashLine(35,50)(65,50){2}
\ArrowLine(90,10)(65,50)
\ArrowLine(65,50)(90,90)
\put(8,20){$\bar{f}'$}
\put(8,70){$f$}
\put(45,55){$\tilde{f}$}
\put(87,20){$\bar{F}'$}
\put(87,70){$F$}
\end{picture}
%
\begin{picture}(100,100)(-12,0)
\ArrowLine(10,90)(50,65)
\ArrowLine(50,65)(90,90)
\DashLine(50,65)(50,35){2}
\ArrowLine(90,10)(50,35)
\ArrowLine(50,35)(10,10)
\put(16,20){$\bar{f}'$}
\put(16,75){$f$}
\put(54,45){$\tilde{f}$}
\put(79,20){$\bar{F}'$}
\put(79,75){$F$}
\end{picture}
%
\begin{picture}(100,100)(-24,0)
\ArrowLine(10,90)(50,65)
\ArrowLine(90,90)(50,65)
\DashLine(50,65)(50,35){2}
\ArrowLine(50,35)(90,10)
\ArrowLine(50,35)(10,10)
\put(16,20){$\bar{f}'$}
\put(16,75){$f$}
\put(54,45){$\tilde{f}$}
\put(79,20){$F$}
\put(79,75){$\bar{F}'$}
\end{picture}
\caption{\label{baba}  Generic Feynman diagrams for 
  \protect$f\bar{f}'\rightarrow F\bar{F}'$
  scattering including $s$- and $t$-channel exchange of $\gamma/Z/W$, 
  and $s$-, $t$- and $u$-channel exchange of sfermion $\tilde{f}$.}
\end{figure}
While the $s$- and $t$-channel $\gamma,Z,W$ amplitudes in the Standard
Model involve the coupling of vector currents, the sfermion exchange
is described by scalar couplings. By performing appropriate Fierz
transformations, the $s$-channel $\ti{f}$ exchange amplitudes can 
be rewritten, however, as $t$-channel vector amplitudes, and
$t/u$-channel $\ti{f}$ exchange amplitudes as $s$-channel vector
amplitudes; for the operators:
\beq 
(\bar{f}_R f'_L)(\bar{F}_L F'_R) \ra
-\frac{1}{2}(\bar{f}_R\gamma_{\mu} F'_R)(\bar{F}_L\gamma_{\mu}f'_L)
\eeq 
The independent $s$-channel amplitudes $f^s_{ij}$ ($i,j=L,R$)
can then be written as follows 
\newcommand{\half}{\frac{1}{2}}
\beq 
f^s_{ij} &=& \frac{Q^s_{ij}}{s} 
 +\half\; \frac{G^t_{ij}/e^2}{t-m^2_{\ti{f}}}
 -\half\; \frac{G^u_{ij}/e^2}{u-m^2_{\ti{f}}} \label{helamps}
\eeq
where $s=(p_f+p_{\bar{f}'})^2$, $\sqrt{s}$ is the center-of-mass
energy of the $f\bar{f}'$ system,
$t=(p_f-p_F)^2=-s(1-\cos\theta)/2$, and
$u=(p_f-p_{\bar{F}'})^2=-s(1+\cos\theta)/2$. 
Note the relative sign between $t$- and $u$-channel sfermion
contributions due to different ordering of fermion operators in
the Wick reduction \cite{lq}. 
Similarly, the $t$-channel exchange amplitudes $f^t_{ij}$ 
read
\begin{eqnarray}
f^t_{ij} &=& \frac{Q^t_{LR}}{t} 
+\half\frac{G^s_{ij}/e^2}{s-m^2_{\ti{f}}+i\Gamma_{\ti{f}} m_{\ti{f}}}
\label{helampt}
\end{eqnarray} 
The parameters $m_{\ti{f}}$ and $\Gamma_{\ti{f}}$ are the mass 
and width of the
exchanged sfermion $\ti{f}$ ($\tilde{f}$ is a generic notation
of the exchanged sfermion, not necessarily the superpartner of $f$).  
To simplify notations we
have defined the indices $L,R$ to denote the helicities of the {\it
incoming fermion $f$} (first index) and the {\it outgoing antifermion $F'$}
(second index).  The helicities of the incoming antifermion and the
outgoing fermion are fixed by the $\gamma_5$ invariance of the vector
interactions: they are opposite to the helicities of the fermionic 
partner in $s$-channel amplitudes and the same in $t$-channel
amplitudes. The generalized SM charges $Q^{s,t}_{ij}$  for gauge
boson exchanges and the factors $G^{s,t,u}_{ij}$ in terms of
Yukawa couplings of the exchanged sfermions will be given when
specific reactions are discussed. Note that the sign of the SM
charges determines the interference pattern of gauge boson with
sfermion exchange terms.

\section{Yukawa Couplings}
The masses and Yukawa couplings of sfermions are not predicted
by theory. 
At energies much lower than the sparticle masses, $R$-parity breaking
interactions can be formulated in terms of effective $llll$ and $llqq$ contact
interactions.  These operators will in general mediate $L$ violating
processes and FCNC processes.  Since the existing data are
consistent with the SM,  stringent
constraints on the Yukawa couplings and masses can be derived \cite{limits}.  
However, if only some of the terms
with a particular generation structure are present in eq.\ 
(\ref{effl}), then the effective four-fermion Lagrangian is not strongly 
constrained.  Similarly, the couplings can be arranged such
that there are no other sources of FCNC interactions than CKM mixing
in the quark sector. Below we will consider the following two 
scenarios:
\\[1mm]
($i$) one single Yukawa coupling is large, all the other couplings
are small and thus neglected; \\[1mm]
($ii$) two Yukawa couplings which violate {\it one and the same}
lepton flavor are large, all the others are neglected.

Since theoretically the third-generation sfermions are expected
lighter than the first two and, due to large top quark
mass, the violation of the third-generation lepton-flavor might
be expected maximal, we will concentrate on possible effects
generated by $\tilde{\tau}$ and $\tilde{\nu}_{\tau}$, $i.e.$ we
are concerned with $\lambda_{i3i}$ and $\lambda'_{3jk}$
couplings.
In these cases low-energy experiments are not restrictive and
typically allow for couplings $\lambda \lsim 0.1\times$($\ti{m}$/200
GeV), where $\ti{m}$ is the mass scale of the sparticles participating
in the process.  The corresponding limits relevant for $\lambda_{i3i}$
and $\lambda'_{3jk}$, derived by assuming only one non-vanishing
coupling at a time, are summarized in Table~\ref{tablam}.
\begin{table}[htbp]
\begin{center}
\footnotesize\rm
\caption{Low-energy limits for the couplings 
\protect$\lambda_{i3i}$
($i=1,2$) and \protect$\lambda'_{3jk}$ ($j=1,2$, $k=1,2,3$)  
assuming the relevant sfermion masses \protect$\ti{m}=200$ GeV. 
They are derived from 
(a) \protect$\Gamma(\tau\ra e\nu\bar{\nu})/ 
   \Gamma(\tau\ra\mu\nu\bar{\nu})$ \protect\cite{snu}; 
(b) \protect$\Gamma(\tau\ra e\nu\bar{\nu})/ 
   \Gamma(\mu\ra e\nu\bar{\nu})$ \protect\cite{barger}; 
(c) $K\ra \pi\nu\nu$ \protect\cite{agashe}; 
(d) $D\bar{D}$ mixing \protect\cite{D1}; 
(e) $\tau\ra \pi\nu$ \protect\cite{deb}.}  
\begin{tabular}{ccccc}\topline
coupling&  $\lambda_{131}$& $\lambda_{232}$& $\lambda'_{3jk}$& 
$\lambda'_{31k}$  \\[1mm]
\midline
Low-energy  & 0.08$^a$& 0.08$^b$ & 0.024$^c$ & 0.32$^e$ \\ 
limit       &         &          & 0.34$^d$  & \\
\bottomline
\end{tabular}
\label{tablam}
\end{center} 
\end{table}
The limit (d) for $\lambda'_{3jk}<0.34$ is derived assuming the CKM
mixing due to absolute mixing in the up-quark sector only ($NDD^c$
diagonal); if the CKM mixing is due to absolute mixing in the
down-quark sector ($EQD^c$ diagonal), more stricter bound (c) of 0.024
applies.  In summary, present low-energy data allow
$\lambda_{i3i}\lsim 0.08$, and $\lambda_{i3i}\lambda'_{311}\lsim
(0.05)^2$, even for the limit (c).

\section{Sleptons at LEP2}
If $\lambda_{131}\ne 0$, the tau sneutrino $\ti{f}=\ti{\nu}_{\tau}$ can
contribute to Bhabha scattering via $s$- and $t$-channel
exchanges, and the electron sneutrino $\ti{f}=\ti{\nu}_e$ in the
process   $e^+e^-\ra\tau^+\tau^-$ can be exchanged in the 
$t$-channel. Assuming in addition $\lambda_{232}\ne0$, also muon
pair production, $e^+e^-\ra\mu^+\mu^-$, can be mediated by the
$s$-channel $\ti{\nu}_{\tau}$ resonance.  
Taking $\lambda'_{3jk}\ne0$ would lead to $s$-channel
$\ti{\nu}_{\tau}$ contribution in hadronic processes $e^+e^-\ra
q_j\bar{q}_k$. 
We will consider these cases below. Note that apart from
$R$-parity violating decays, the $\ti{\nu}_{\tau}$ can also
decay via $R$-parity conserving modes; such decays have already been
discussed in the literature \cite{dl}. On the other hand,
$\ti{\tau}$ slepton in $e^+e^-$ collisions can only contribute
via $t/u$-channels to the neutrino-pair production cross section, which for
couplings considered here is below 1 \%.  \\[1mm]
\noindent (a) {\it Bhabha scattering:} 
In this case the differential cross section is given by
eq.~(\ref{dsigdcos}) with $A_c=1$. The SM generalized charges in
helicity amplitudes are as follows
\begin{eqnarray}
&&Q^s_{ij} = 1+ g^e_i
g^e_{-j}\frac{s}{s-m^2_Z+i\Gamma_Zm_Z}\label{opposite} \\
&&Q^t_{ij} = 1+ g^e_i g^e_{-j}\frac{t}{t-m^2_Z} \non
\end{eqnarray}
The subscript $-j$ means that the helicity index is opposite to
$j$ because in eqs.~(\ref{helamps},\ref{helampt}) the outgoing
positron with the 
helicity $L(R)$ couples with the charge   $g_R (g_L)$.  
 The left/right
$Z$ charges  of the fermions are defined as
\begin{eqnarray}
g^f_L=(\frac{\sqrt{2}G_{\mu}m^2_Z}{\pi\alpha})^{1/2}
(I_3^f-s^2_W Q^f), \mbox{~~~~~~~} 
g^f_R=(\frac{\sqrt{2}G_{\mu}m^2_Z}{\pi\alpha})^{1/2}
(  {} -s^2_W Q^f)\non 
\end{eqnarray}
The sneutrino contributions are specified in terms of the 
factors $G_{ij}$  as follows
\begin{eqnarray}
G^s_{LL}=G^s_{RR}=G^t_{LL}=G^t_{RR}=(\lambda_{131})^2
\end{eqnarray}
with all other $G_{ij}=0$. Note that the $s$-channel ($t$-) sneutrino
exchange interferes with the $t$-channel ($s$-) $\gamma,Z$
exchanges. 
\\[1mm]
\noindent (b) {\it Muon-pair production:} 
Since the $t$-channel $\gamma,Z$ and $\ti{\nu}_{\tau}$ exchanges
are absent, $Q^t_{ij}=0$, $G^t_{ij}=0$, the $s$-channel
sneutrino exchange does not interfere with the SM processes. The
SM generalized charges $Q^s_{ij}$ are given by
eq.~(\ref{opposite}), and the sneutrino process gives  
\begin{eqnarray}
G^s_{LL}=G^s_{RR}=\lambda_{131}\lambda_{232}, 
\mbox{~~~~all~other~~~}G_{ij}=0
\end{eqnarray}
\noindent (c) {\it Tau-pair production:} 
In the scenario considered here, this process can receive only
the $t$-channel exchange of $\ti{\nu}_e$ with 
\begin{eqnarray}
G^t_{RR}=(\lambda_{131})^2, \mbox{~~~~all~other~~~} G_{ij}=0 
\end{eqnarray}  
which will interfere with the SM $\gamma,Z$ $s$-channel
processes with $Q^s_{ij}$ given by eq.~(\ref{opposite}).\\[1mm] 
\noindent (d) {\it $e^+e^-$ annihilation to hadrons:}
For the down-type quark-pair production, $e^+e^-\ra d_k\bar{d}_k$,
the differential cross section is also given by
eq.~(\ref{dsigdcos}), however with $A_c=3$.  
In this case the situation
is similar to the muon-pair production
process: there is no interference between $s$-channel
$\ti{\nu}_{\tau}$ exchange, given by 
\begin{eqnarray}
G^s_{LL}=G^s_{RR}=\lambda_{131}\lambda'_{3kk}
\end{eqnarray}
and the SM $\gamma,Z$ processes, with the generalized charges
\begin{eqnarray}
Q^s_{ij}=-Q^q +g_i^eg_{-j}^q \frac{s}{s-m^2_Z+i\Gamma_Zm_Z}
\end{eqnarray}
while all other $Q_{ij}$ and $G_{ij}$ vanish. 
The up-type quark-pair production is not affected by sneutrino
processes, as can be easily seen from the general structure of
couplings in eq.~(\ref{effl}). Finally, the unequal-flavor
down-type quark-pair production process, $e^+e^-\ra
d_j\bar{d}_k$, can be generated only by $s$-channel sneutrino with 
$G^s_{LL}=G^s_{RR}=\lambda_{131}\lambda'_{3jk}$.

\begin{figure}[htbp] 
\unitlength 1mm
\begin{picture}(80,150)
  \put(-10,25){ \epsfxsize=13cm \epsfysize=14cm \epsfbox{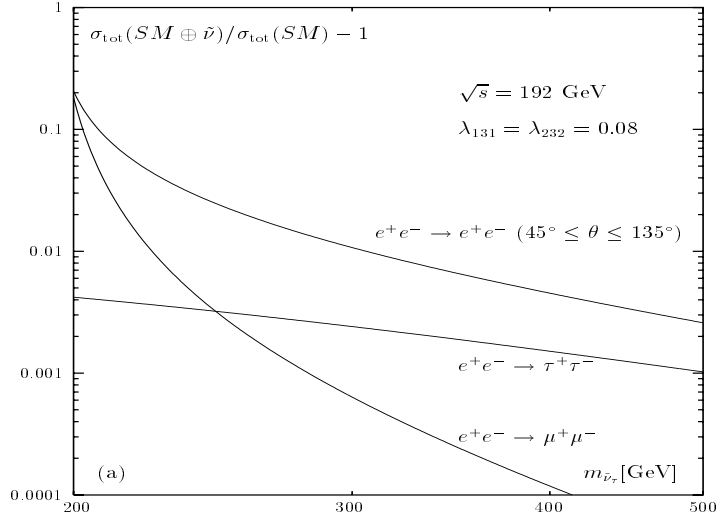}}
  \put(-10,-50){ \epsfxsize=13cm \epsfysize=14cm
\epsfbox{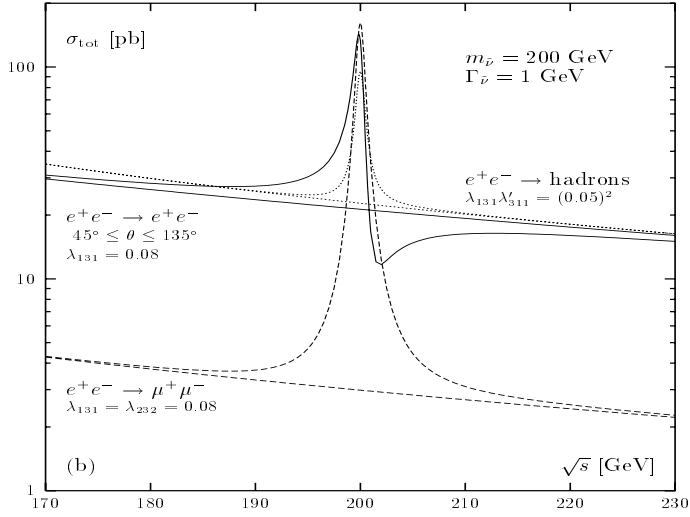}}
\end{picture}
\caption{(a)
  Effect of sneutrino $\ti{\nu}_{\tau}$ exchange on the cross section
  for Bhabha scattering (restricting $45^{\circ} \leq \theta \leq
  135^{\circ}$), and $\mu^+\mu^-$ and $\tau^+\tau^-$ production at
  $\protect\sqrt{s}=192$ GeV. (b) Cross section for Bhabha scattering
  (solid lines), $\mu^+\mu^-$ (dashed lines) and hadron production
  (dotted lines) in the SM, and including $\ti{\nu}_{\tau}$ sneutrino
  resonance formation as a function of the $e^+e^-$ energy. }
\label{figlep}
\end{figure}

The impact of sneutrino $\ti{\nu}_{\tau}$ exchange on processes (a--c)
at LEP2 energy is shown in Fig.~2a, where the change of the total
cross section, $\sigma_{\rm tot}(SM\oplus \tilde{\nu})\ /\sigma_{\rm
  tot}(SM)-1$, is plotted as a function of sneutrino mass assuming
$\lambda_{i3i}=0.08$ and $\lambda_{131}\lambda'_{3jk}=(0.05)^2$. For
Bhabha scattering the scattering angle is restricted to $45^{\circ}
\leq \theta \leq 135^{\circ}$. In processes (a) and (b), where the
$s$-channel sneutrino exchange can contribute, the effect can be very
large for sneutrino mass close to the LEP2 center-of-mass energy. Note
the difference due to different interference pattern between Bhabha
scattering and tau-pair production on the one hand, and muon-pair
production processes on the other: Bhabha and tau-production processes
are more sensitive to heavy sneutrinos.  If sneutrino is within the
reach of LEP2, a spectacular resonance can be observed in Bhabha
scattering, muon-pair, and/or quark-pair production processes, Fig.2b;
again different interference patterns are seen. In the calculations
the total decay width $\Gamma_{\ti{\nu}_{\tau}}=1$ GeV has been
assumed.  The partial decay width $\Gamma(\ti{\nu}_{\tau}\ra
e^+e^-)=\lambda_{131} m_{\ti{\nu}_{\tau}}/16\pi$ is very small.
However, sneutrinos can also decay via $R$-parity conserving couplings
to $\nu\chi^0$ and $\ell^\pm\chi^\mp$ pairs with subsequent $\chi^0$
and $\chi^\pm$ decays. The partial decay widths into these channels
depend on the choice of supersymmetry breaking parameters. In large
regions of the parameter space, the total decay width is found to be
as large as 1 GeV, which is significantly larger than the energy
spread $\delta E \sim 200$ MeV at LEP2.  In this case the interference
with the SM processes must be taken into account. The peak cross
section for Bhabha scattering is given by the unitarity limit
$\sigma_{peak}=8\pi B_e^2/m^2_{\ti{\nu}_{\tau}}$ with sneutrino and
antisneutrino production added up, where $B_e$ is the branching ration
for the sneutrino decay to $e^+e^-$. An interesting situation may
occur if sneutrinos mix and mass eigenstates are split by a few GeV
\cite{snumix}.  Then one may expect two separated peaks with reduced
maximum cross sections to observe in the energy dependence in Fig.2b
for the processes (a), (b) and/or (d).

The angular distribution of leptons and quark jets is nearly isotropic
on the sneutrino resonance. As a result, the strong forward-backward
asymmetry in the Standard Model continuum is reduced to $\sim 0.03$ on
top of the sneutrino resonance. The deviations of the Bhabha cross
section from the SM expectations would allow to determine directly the
$\lambda_{131}$ coupling, or to derive an upper limit.  Similarly from
the other processes one could derive limits for $\lambda_{232}$ and
$\lambda'_{3jk}$. For example, if the total hadronic cross section at
192 GeV can be measured to an accuracy of 1\%, the Yukawa couplings
for a 200 GeV sneutrino can be bounded to
$\lambda_{131}\lambda'_{311}\lsim (0.045)^2$ \cite{tev}.  Recently preliminary
  results for some of the couplings from LEP 172 GeV data have been
  published \cite{lepres}.

\section{Sleptons at Tevatron}
At the Tevatron the case $\lambda'_{311}$ is the most
interesting since it allows for $\ti{\nu}_{\tau}$ and $\ti{\tau}$ 
resonance formation in valence quark collisions. 
Even though the sneutrinos and charged sleptons are expected to
have small widths ($\sim 1$ GeV or less), it will be difficult
to detect their decay to quarks in the hadronic environment.
Therefore we will consider leptonic decays of sleptons via
$\lambda_{i3i}$ couplings. To be specific we will consider
$\lambda_{131}$ and discuss $e^+e^-$ and $e^+\nu_e$ production
in $p\bar{p}$ collisions; the same results hold for $\mu^+\mu^-$
and $\mu^+\nu_{\mu}$ production if $\lambda_{232}$ is
assumed.

For $p\bar{p}\ra e^+e^-$ and $e^+\nu_e$ the differential cross
sections are obtained by combining the parton cross sections
with the luminosity spectra for quark-antiquark annihilation
\begin{eqnarray}
\frac{\mbox{d}^2\sigma}{\mbox{d}M_{\ell\ell}\mbox{d}y} 
[p\bar{p}\ra \ell_1\ell_2]=\sum_{ij} \frac{1}{1+\delta_{ij}}
\, \left(f_{i/p}(x_1)  f_{j/\bar{p}}(x_2) +(i 
\leftrightarrow j)\right)\,  \hat{\sigma}  
\label{lum}
\end{eqnarray}
where $\hat{\sigma}$ is the cross section for the partonic subprocess
$ij\ra\ell_1\ell_2$, $\ell_1\ell_2=e^+e^-$ or $e^+\nu$,
$x_1=\sqrt{\tau}e^y$, $x_2=\sqrt{\tau}e^{-y}$.  $M_{\ell\ell}=(\tau
s)^{1/2} = (\hat{s})^{1/2}$ is the mass and $y$ the rapidity of the
lepton pair.  The probability to find a parton $i$ with momentum
fraction $x_i$ in the (anti)proton is denoted by
$f_{i/p(\bar{p})}(x_i)$.

The partonic differential cross sections in the
$q\bar{q}^{(')}$ center-of-mass 
frame are given by eq.~(\ref{dsigdcos}) with $A_c=1/3$, and $s$,
$t$ and $u$ replaced by $\hat{s}$, $\hat{t}$ and $\hat{u}$ which
refer to the $q\bar{q}^{(')}\ra \ell_1\ell_2$ subprocess. 
The $e^+e^-$ and $e^+\nu_e$ production processes are specified
as follows\\[1mm]
\noindent (a) {\it The process $q\bar{q}\ra e^+e^-$:} 
The $s$-channel sneutrino $\ti{\nu}_{\tau}$ exchange contributes
only to $d\bar{d}$ scattering with 
\begin{eqnarray}
G^s_{LL}=G^s_{RR}=\lambda_{131}\lambda'_{311}
\end{eqnarray}
which does not interfere with  the SM $s$-channel $\gamma,Z$
processes, for which the generalized charges are as follows
\begin{eqnarray}
Q^s_{ij}=-Q^q+ g^q_L g^e_{-j}\frac{\hat{s}}{\hat{s}-m^2_Z
+i\Gamma_Z m_Z}
\end{eqnarray}
All other $Q_{ij}$ and $G_{ij}$ vanish.\\[1mm]
\noindent (b) {\it The process $u\bar{d}\ra e^+\nu_e$:} 
This process proceeds via the $s$-channel $W$ boson and
$s$-channel $\ti{\tau}$ slepton exchanges. Only
\begin{eqnarray}
&&Q^s_{LR}=\frac{1}{2\sin^2\theta_W}\,\frac{\hat{s}}
{\hat{s}-m^2_W+i\Gamma_Wm_W}\\
&&G^s_{LL} =-\lambda_{131}\lambda'_{311}
\end{eqnarray}
are non-zero; all other $Q_{ij}$ and $G_{ij}$
vanish. 

\begin{figure}[htbp] 
\unitlength 1mm
\begin{picture}(80,150)
  \put(-10,25){ \epsfxsize=13cm \epsfysize=14cm \epsfbox{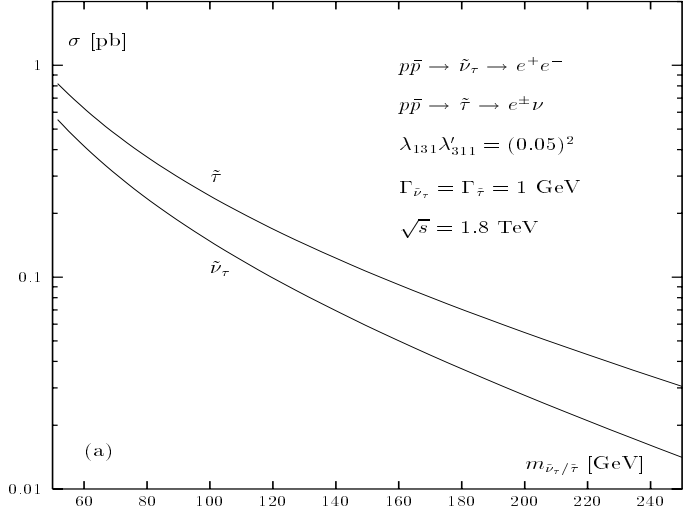}}
  \put(-10,-50){ \epsfxsize=13cm \epsfysize=14cm \epsfbox{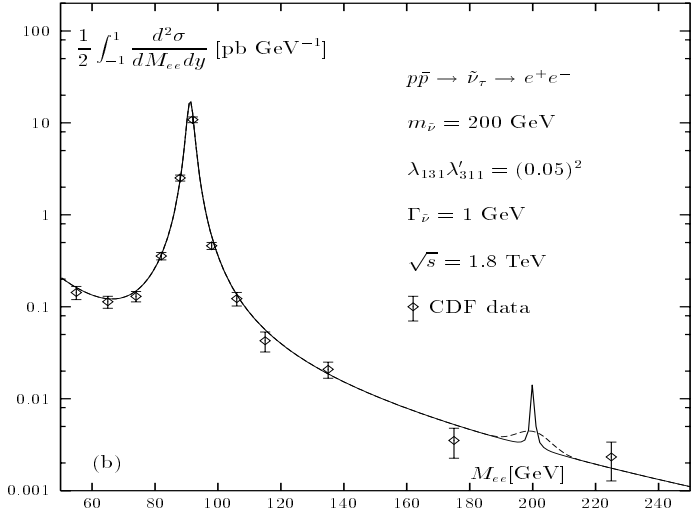}}
\end{picture}
\caption{(a) The cross section for sneutrino and antisneutrino 
  ($\ti{\nu}_{\tau}$) and stau ($\ti{\tau}$) production at the
  Tevatron, including the branching ratios to lepton-pair decays.  (b)
  The $e^+e^-$ invariant mass distribution including the $s$-channel
  sneutrino in the channel $d\bar{d}\ra e^+e^-$ is compared with the
  CDF data; solid line: ideal detector, dashed line: sneutrino
  resonance smeared by a Gaussian width 5 GeV. The CTEQ3L structure
  functions have been used. }
\label{figtev}
\end{figure}

The total cross sections for sneutrino and charged slepton production
in $e^+e^-$ and $e^+\nu_e$ channels, respectively, at Tevatron are
shown in Fig.3a as a function of slepton mass. The total decay widths
of sleptons have been set to a typical value of 1 GeV, corresponding
to the branching ratios for leptonic decays of order 1\%. The
di-electron invariant mass distribution is compared to the CDF data in
Fig.3b, where, following CDF procedure \cite{CDF}, the prediction for
$\frac{1}{2}\int^{1}_{-1}
{\mbox{d}^2\sigma}/{\mbox{d}M_{ee}\mbox{d}y}$ is shown. The
solid line corresponds to an ideal detector, while the dashed curve
demonstrates the distribution after the smearing of the peak by
experimental resolution characterized by a Gaussian width of 5 GeV. In
both plots the CTEQ3L parametrization \cite{cteq} is used together
with a multiplicative $K$ factor for higher order QCD corrections to
the SM Drell-Yan pair production.  The corresponding $K$ factor for
slepton production has not been calculated yet, leading to a
theoretical uncertainty in the $\lambda\lambda'$ couplings at a level
of about 10\%.
Assuming the sneutrino contribution to be smaller than the
experimental errors, we estimate that the bound
$\lambda_{131}\lambda'_{311}\lsim (0.08)^2 \ti{\Gamma}^{1/2}$ can be
established \cite{tev}, where $\ti{\Gamma}$ denotes the sneutrino width in
units of GeV.

\section{Summary}
The $R$-parity violating formulation of MSSM offers a distinct
phenomenology. If the lepton-flavor violating couplings are close to
current low-energy limits, and the slepton masses are in the range of 200
GeV, spectacular events can be expected at both LEP2 and Tevatron.  On
the other hand, if no deviations from the SM expectations are
observed, stringent bounds on individual couplings can be derived
experimentally in a direct way.

\section*{Acknowledgments}
It is a pleasure to thank H.V. Klapdor-Kleingrothaus and H.~Paes
for their kind invitation and warm hospitality at the workshop.
I am grateful to R.~R\"uckl, H.~Spiesberger and P.~Zerwas
for their collaboration and comments on the manuscript. I would also
like to thank G.~Bhattacharyya, J.~Conway, S.~Eno, H.~Dreiner, 
W.~Krasny, K.~Maeshima, W.~Sakumoto, Y.~Sirois and F.~\.Zarnecki for
discussions and communications.

\end{document}